\documentclass[conference, a4paper]{IEEEtran}
\IEEEoverridecommandlockouts
\usepackage{cite}
\usepackage{amsmath,amssymb,amsfonts}
\usepackage{algorithmic}
\usepackage{graphicx}
\usepackage{textcomp}
\usepackage{xcolor}
\usepackage{enumitem}
\usepackage{tabularray}
\usepackage{color,soul}
\def\BibTeX{{\rm B\kern-.05em{\sc i\kern-.025em b}\kern-.08em
    T\kern-.1667em\lower.7ex\hbox{E}\kern-.125emX}}
\begin{document}

\title{Effective Fine-Tuning of Vision Transformers with Low-Rank Adaptation for Privacy-Preserving Image Classification 
\\
\thanks{This work was supported by JSPS KAKENHI Grant Number 25K07750.}
}

\author{\IEEEauthorblockN{Haiwei Lin}
\IEEEauthorblockA{\textit{Graduate School of Informatics} \\
\textit{Chiba University}\\
Chiba, Japan \\
25rd1008@student.gs.chiba-u.jp}
\and
\IEEEauthorblockN{Shoko Imaizumi}
\IEEEauthorblockA{\textit{Graduate School of Informatics} \\
\textit{Chiba University}\\
Chiba, Japan \\
imaizumi@chiba-u.jp}
\and
\IEEEauthorblockN{Hitoshi Kiya}
\IEEEauthorblockA{\textit{Faculty of System Design} \\
\textit{Tokyo Metropolitan University}\\
Tokyo, Japan \\
kiya@tmu.ac.jp}
}

\maketitle

\begin{abstract}
We propose a low-rank adaptation method for training privacy-preserving vision transformer (ViT) models \textcolor{black}{that} efficiently freezes pre-trained ViT model weights. In the proposed \textcolor{black}{method}, trainable rank decomposition matrices are injected into each layer of the ViT architecture\textcolor{black}{,} and moreover\textcolor{black}{,} the patch embedding layer is not frozen, \textcolor{black}{unlike in the case of the} conventional low-rank adaptation methods. The \textcolor{black}{proposed} method allows us not only to reduce the number of trainable parameters but to also maintain almost the same accuracy as that of full-time tuning. 

\end{abstract}


\section{Introduction}
The importance of \textcolor{black}{vision transformer} (ViT) based-models \cite{ViT} has been increasing in recent years. 
ViT-based models can be applied to vision-language tasks \cite{A_10445007} in addition to image classification, object detection \cite{B_beal2020toward}, and semantic segmentation tasks \cite{C_Zheng_2021_CVPR}. Models based on transformer architectures involve fine-tuning from weights pretrained with the ImageNet dataset in general. However, with recent advances in \textcolor{black}{large-scale} pre-training, full-time tuning, which retrains all model parameters, becomes less feasible due to limited storage space and time-consuming model training. To overcome the problem, low-rank adaptation methods have been studied so far. LoRA \cite{LoRA} was proposed as a low-rank adaptation method \textcolor{black}{for} large language models. LoRA was extended to MeLo \cite{melo} so that it can be applied to ViT. 

\textcolor{black}{ViT-based Models trained with perceptually encrypted data} have been studied for privacy-preservation \cite{D_kiya2022overview, E_8486525, F_9247223}, access control \cite{G_maungmaung2021protection} and adversarial defenses \cite{H_9190904, I_10530249}. However, it has not been confirmed whether existing low-rank methods are effective in training models with encrypted data.

Accordingly, in this paper, we point out that MeLo is not effective in training models with perceptually encrypted data. In addition, we propose a novel low-rank adaptation method for training ViT-based models.

\section{Proposed Method}
\begin{figure}[t]
    \centering
    \includegraphics[width=1\linewidth]{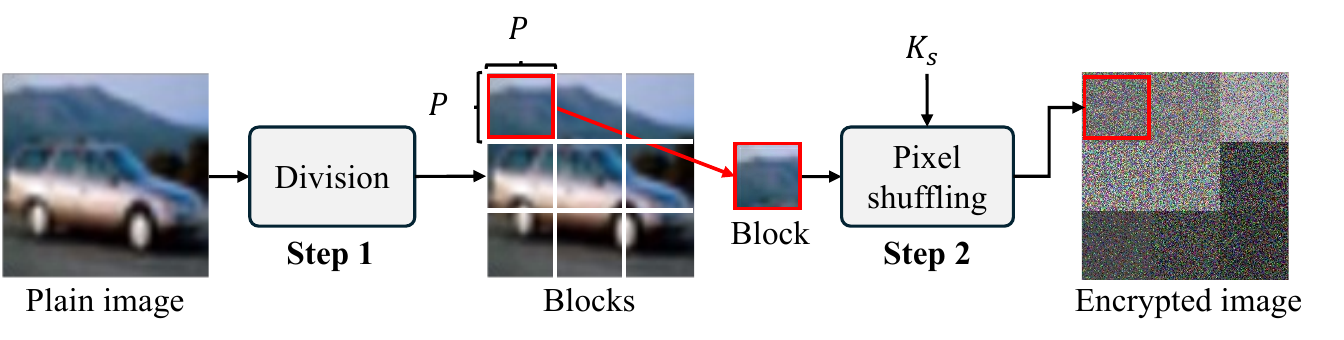}
    \caption{Pipeline of block-wise image encryption}
    \label{fig:encryption}
\end{figure}
\subsection{Block-wise image encryption}\label{sec:2_1}
We \textcolor{black}{use} a block-wise image encryption method to protect sensitive visual information \textcolor{black}{in training} and test images.
As shown in Figure \ref{fig:encryption}, the encryption process of the proposed method can be summarized as follows.
\begin{enumerate}[label=\textbf{Step \arabic*:}, align=left, leftmargin=*, labelsep=1em]
\item Divide an image into non-overlapping blocks with the same size as the patch size $P \times P$ of a pretrained ViT.
\item Randomly shuffle pixels across three color channels within each block by using a single random sequence generated with a key $K_s$.

\end{enumerate}
The above procedure is carried out in the same manner as in \cite{DomainAdaptation}. Throughout all training and test images, every block is constantly encrypted with the identical key $K_s$.

\textcolor{black}{
\subsection{Fine-tuning using MeLo}
\begin{figure*}[t]
    \centering
    \includegraphics[width=1.0\linewidth]{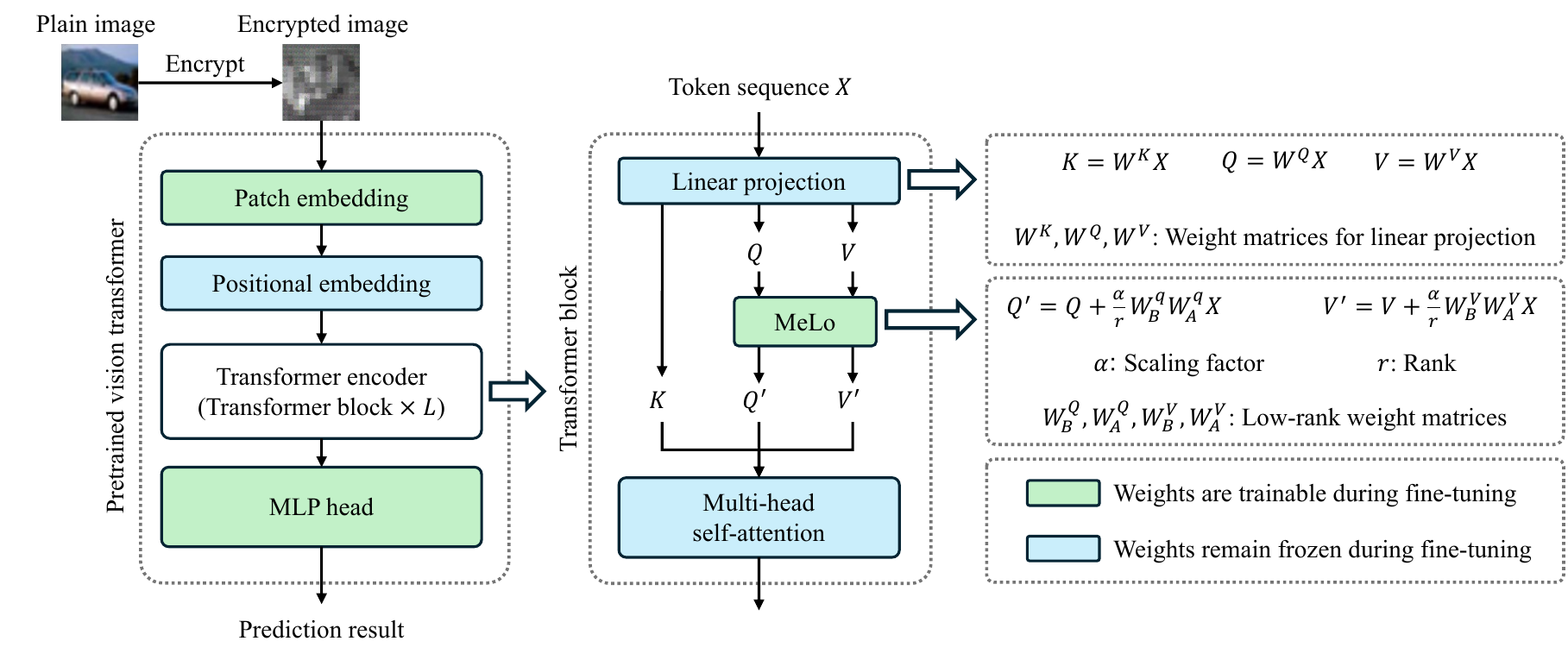}
    \caption{Overview of proposed fine-tuning method.}
    \label{fig:overview}
\end{figure*}
Full-time tuning is usually used to fine-tune a pre-trained model. In contrast, to reduce the number of trainable parameters, various low-rank adaptation methods have been proposed. 
For MeLo (Medical image Low-rank adaptation) \cite{melo}, even when using only a few trainable parameters, competitive results can be achieved by fixing some \textcolor{black}{of the} weights of ViT models and only adding small low-rank plug-ins. However, existing low-\textcolor{black}{rank} methods including MeLo have not considered the influence of encrypted data.
}

\subsection{Low-rank adaptation for privacy preserving learning}
To reduce the influence of image encryption, we propose a novel low-rank adaptation method \textcolor{black}{that} is an extension of MeLo.
The main idea of MeLo is to freeze many pre-trained model weights and inject trainable rank decomposition matrices, which are the pre-trained query and value projection matrices denoted as $W^Q$ and $W^V$, into each layer of the \textcolor{black}{transformer} architecture for fine-tuning as shown in Figure \ref{fig:overview}.

Given a token sequence $X \in \mathbb{R}^{d \times l}$, it is linearly projected into three vector representations: $Q, K, V \in \mathbb{R}^{d \times l}$, where $d$ is the embedding dimension, and $l$ is the length of $X$. These vector representations are computed as $K=W^KX$, $Q=W^QX$\textcolor{black}{,} and $V=W^VX$, where $W^Q, W^K$, \textcolor{black}{and} $W^V \in \mathbb{R}^{d \times d}$ are weight matrices used for linear projection. After the injection of MeLo weights, trainable low-rank weight matrices are added to the computations of $Q$ and $V$ as
\begin{equation}
    Q^{'} = Q +  \frac{\alpha}{r}W_B^QW_A^QX,
\end{equation}
\begin{equation} 
    V^{'} = V +  \frac{\alpha}{r}W_B^VW_A^VX.
\end{equation}
\textcolor{black}{
Here, $\alpha$ represents a scaling factor\textcolor{black}{,} and $r$ is the rank. $W_B^Q$ and $W_B^Q \in \mathbb{R}^{d \times r}$, $W_A^Q$, and $W_A^Q \in \mathbb{R}^{r \times d}$ are low-rank weights. Note that $r$ satisfies the condition $r \ll d$.
}

\textcolor{black}{
However, MeLo freezes weights in the patch and position embedding. Such constraints \textcolor{black}{make it} difficult to protect against the influence of image encryption because perceptual image encryption is commonly carried out on the basis of the permutation operation of pixel values. 
Accordingly, \textcolor{black}{our} novel method does not freeze patch embedding in addition \textcolor{black}{to using} MeLo weights. 
In experiments, it will be confirmed that the use of encrypted images \textcolor{black}{affects} patch embedding in terms of image classification accuracy and the number of trainable parameters.
}

\begin{figure}[t]
    \centering
    \includegraphics[width=1.0\linewidth]{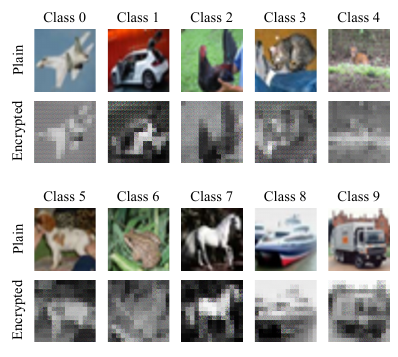}
    \caption{Samples of plain images and their encrypted images from each class of CIFAR-10.}
    \label{fig:sample}
\end{figure}
\section{Experiment}
We conducted image classification experiments on the CIFAR-10 dataset and \textcolor{black}{a} ViT pretrained on ImageNet-1K. Both training and test images were enlarged to $224 \times 224$ pixels\textcolor{black}{,} and the images were \textcolor{black}{then} encrypted with the same key.
The block size for block-wise encryption was set to $16 \times 16$\textcolor{black}{, which was} identical to the patch size of the pretrained ViT. 
Samples of plain images and their encrypted images from each class of CIFAR-10 are shown in Figure \ref{fig:sample}. \textcolor{black}{MeLo} weights were configured with a scaling factor $\alpha = 4$ and rank $r = 8$. In fine-tuning, we used a learning rate of $1 \times 10^{-4}$ and \textcolor{black}{the} Adam optimizer to train the ViT for $200$ epochs. We also conducted experiments with \textcolor{black} {the} original MeLo, \textcolor{black}{in} which \textcolor{black}{only the} MeLo weights \textcolor{black}{were updated}, and full fine-tuning.

\textcolor{black}{The} experimental results are summarized in Table \ref{tab:result}, where Acc (Plain) denotes the accuracy of models fine-tuned with plain images, while Acc (Encrypted) denotes the accuracy of models fine-tuned with encrypted images. Although full fine-tuning showed favorable accuracies on both plain and encrypted images, it required a large number of trainable parameters for domain adaptation. In comparison, MeLo had the fewest trainable parameters, but its performance on encrypted images was inferior to both full fine-tuning and our method. Our method outperformed full fine-tuning on encrypted images while requiring only $0.71$ million trainable parameters, which was approximately $0.8\%$ of the full fine-tuning parameter count.
\begin{table}[t]
\centering
\caption{Experimental results}
\label{tab:result}
\begin{tblr}{
  cell{1}{1} = {c},
  cell{2}{2} = {c},
  cell{2}{3} = {c},
  cell{2}{4} = {c},
  cell{3}{2} = {c},
  cell{3}{3} = {c},
  cell{3}{4} = {c},
  cell{4}{2} = {c},
  cell{4}{3} = {c},
  cell{4}{4} = {c},
  hline{1,5} = {-}{0.08em},
  hline{2} = {-}{},
}
 & \# of Parameters & Acc (Plain) & Acc (Encrypted)\\
Full fine-tuning & 82.56M & 98.16\% & 96.16\%\\
MeLo \cite{melo} & 0.15M & 98.36\% & 90.05\%\\
Ours & \textbf{0.71M} & 97.98\% & \textbf{96.35\%}
\end{tblr}
\end{table}

\section{Conclusion}
We proposed a novel fine-tuning method using a low-rank adaptation \textcolor{black}{method} that enables privacy-preserving image classification. In experiments, the method was demonstrated to be effective in terms of both the number of trainable parameters and the accuracy of image classification, compared with full fine-tuning and MeLo.

\section{ACKNOWLEDGEMENT}
This work was supported in part by JSPS KAKENHI (Grant Number 25K07750).

\bibliographystyle{ieeetr}
\bibliography{bibliography}


\end{document}